\documentclass{article}
\usepackage{amsmath,amsfonts,bm}

\def\1{\bm{1}}

\def\sp{space}

\def\rvx{{\mathbf{x}}}
\def\rvy{{\mathbf{y}}}

\DeclareMathAlphabet{\mathsfit}{\encodingdefault}{\sfdefault}{m}{sl}
\SetMathAlphabet{\mathsfit}{bold}{\encodingdefault}{\sfdefault}{bx}{n}

\def\gP{{\mathcal{P}}}

\def\gY{{\mathcal{Y}}}

\newcommand{\vsp}{\csname v\sp \endcsname}
\newcommand{\hsp}{\csname h\sp \endcsname}

\DeclareMathOperator*{\argmax}{arg\,max}

\usepackage{spconf,amsmath,graphicx}
\usepackage{amssymb}
\usepackage[sort,numbers]{natbib}
\usepackage{symbols}
\usepackage[usenames]{color}                                                        
\definecolor{mycitecolor}{rgb}{0,0.08,0.45} % ICML Blue 
\usepackage[colorlinks=true, citecolor=mycitecolor]{hyperref}

\newcommand*{\ShowNotes}{} %Exist then show notes.
\definecolor{darkred}{rgb}{0.7,0.1,0.1}
\definecolor{darkgreen}{rgb}{0.1,0.7,0.1}
\definecolor{cyan}{rgb}{0.7,0.0,0.7}
\definecolor{dblue}{rgb}{0.2,0.2,0.8}
\definecolor{maroon}{rgb}{0.76,.13,.28}
\definecolor{burntorange}{rgb}{0.81,.33,0}
\definecolor{tealblue}{rgb}{0.212,0.459, 0.533}
\ifdefined\ShowNotes
  \newcommand{\colornote}[3]{{\color{#1}\bf{#2: #3}\normalfont}}
\else
  \newcommand{\colornote}[3]{}
\fi

\title{Multi-Decoder DPRNN: High Accuracy Source Counting and Separation
}
\name{Junzhe Zhu, Raymond A. Yeh, Mark Hasegawa-Johnson} %\thanks{Thanks to XYZ agency for funding.}}
\address{University of Illinois at Urbana-Champaign}
\begin{document}
%\ninept
%
\maketitle
%
%!TEX root = ../main.tex

\begin{abstract}
We propose an end-to-end trainable approach to single-channel speech separation with unknown number of speakers. Our approach extends the MulCat source separation backbone with additional output heads: a count-head to infer the number of speakers, and decoder-heads for reconstructing the original signals. Beyond the model, we also propose a metric on how to evaluate source separation with variable number of speakers. Specifically, we cleared up the issue on how to evaluate the quality when the ground-truth has \textit{more or less speakers} than the ones predicted by the model. We evaluate our approach on the WSJ0-mix datasets, with mixtures up to five speakers. We demonstrate that our approach outperforms state-of-the-art in counting the number of speakers and remains competitive in quality of reconstructed signals.

%\mh{Comment}
%\ray{Comment}
%\jz{Comment}
\end{abstract}

%!TEX root = ../main.tex

\section{Introduction}
% What is the task and why is it important?
Source separation is the task of decomposing a mixed signal into the original signals prior to the mixing procedure. This is an important task with many downstream applications, \eg, improve the accuracy of automatic speech recognition with multiple speakers~\cite{marti2012automatic}, or separating out singing voices and music~\cite{huang2012singing}. Due to the recent progress in deep learning, supervised methods have received a lot of interest~\cite{huang2014deep, hershey2016deep, wang2018supervised, luo2018tasnet, luo2020dual}. These works formulate the source separation as a regression problem, \ie, given the mixed signal regress the individual components. Various specialized deep-net architectures and losses have been proposed.
For example, ~\cite{kolbaek2017multitalker} proposed a loss which is permutation invariant in the ordering of the speakers, or~\cite{luo2020dual} presented a dual-path RNN architecture to better capture both short and long-term features.
However, these works have focused on the setting where the number of speakers is {\em a priori} known. 

Recently, several works have also considered the case with variable number of speakers. For example, ~\cite{nachmani2020} have proposed a method for separating variable number of speakers, where they train a different model \textit{for every} number of speakers. At test time, they run an activity detector on the largest speaker model to determine to number of speakers and then run the corresponding model for source separation. Another work is~\cite{Takahashi2019RecursiveSS} where they have proposed to iteratively separate out one speaker at a time.
While straight forward, these methods either require training multiple deep-nets or running multiple forward-passes at test-time, both of which scale linearly with the possible number of speakers.

%\ray{TODO: some other work and say why don't need to compare.}

To tackle the aforementioned issues, we propose to train a \textit{single model} with multiple output heads: a count-head to infer the number of speakers, and multiple decoder heads to separate the signals. These output heads share the same backbone feature extractor. Therefore, our method requires a single pass through the network at test time and can be trained from end-to-end. Additionally, we propose a new  metric for evaluating the separation of a variable number of speakers. In particular, our metric considers how to evaluate the quality of the reconstruction when the number of speakers differs between prediction and ground-truth. 

We evaluate our approach on WSJ0-mix dataset, with up to five speaker mixtures. Our approach surpasses all existing approaches in terms of source counting and achieves similar performance to state-of-the-art models in source separation.

% What other people have done

% What we have done

% Results

% Summary

%Single source voice separation have recently made significant progress~\cite{hershey2016deep, luo2020dual, luo2020end}. \ray{TODO: more citations.}

%With largest data-sets and models, 

%% Related works 
%{\noindent \bf Audio source separation.}
%\cite{luo2019conv}
%\jz{TODO}

%{\noindent \bf Network architectures.}
%\cite{bai2019deep}
%\ray{TODO}

%% Some conclusion
%In summary, we

%\input{sections/rel}
%!TEX root = ../main.tex
\begin{figure*}[t]
\centering
\includegraphics[width=0.9\linewidth]{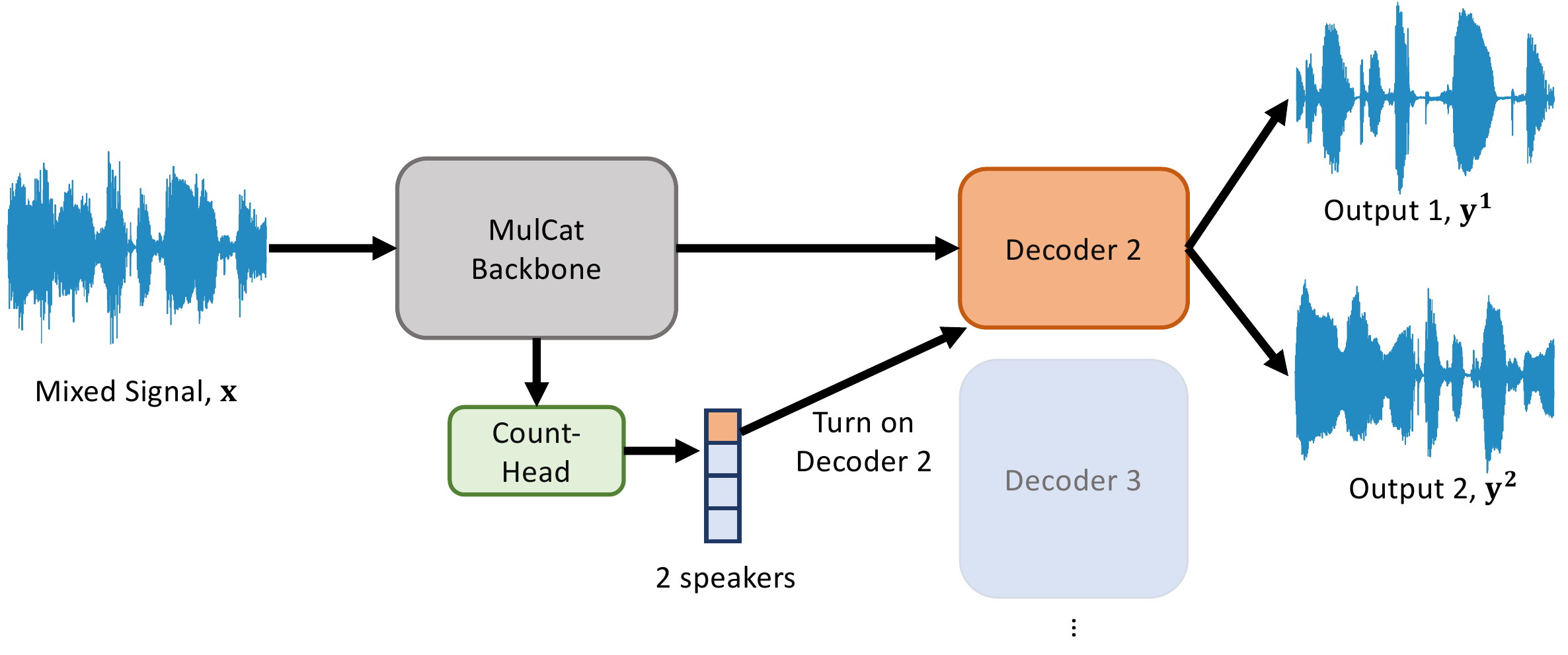}
\vspace{-0.34cm}
\caption{An overview of our proposed approach for handling variable number of speakers for source separation. Given a mixed signal $\rvx$, our model predicts the number of speakers from the mixed signal and uses the corresponding decoder-head to separate the signal. In this case, decoder 2 is selected, hence a reconstruction of two speakers.}
\label{fig:pipeline}
\end{figure*}

\section{Approach}
%A high-level overview of our method is illustrated in. 
We present a single model approach to source separation with a variable number of speakers, illustrated in~\figref{fig:pipeline}. In particular, we augment the standard source separation backbone with additional count-head and decoder-heads to support prediction of variable number of speakers in a single pass. In the following, we describe our approach in more detail.

%{\noindent \bf Problem Formulation:}
\subsection{Problem Formulation}
Let $\rvx$ denote the mixed input audio, and $\gY = \{\rvy^{n}\}$ denote the set of separated audios  from each speaker. The goal is to learn a parametric function,
\be
F_\theta(\rvx) \mapsto \gY,
\ee with trainable parameters $\theta$. One of the challenges is how to construct a model to handle variable number of outputs? For example, a standard deep-net has a fixed number of output dimensions and does not change between examples.

To mitigate this problem, we assume that the maximum number of speakers, $K \geq |\gY|\;\forall (\rvx, \gY) \in \cD$, is known. 
In this case, we can model a deep-net to count the number of speakers and model a decoder-head for \textit{each} number of speakers. This allow us to dynamically select which decoder-head to run and output the correct number of speakers. 

%In more details, 
We propose a single end-to-end trainable deep-net to accomplish this. Our deep-net contains a count-head, which counts the number of speakers in the mixed-audio, and a list of decoder-heads to reconstruct audios for the corresponding number of speakers. These heads share input features extracted from a backbone network~\cite{nachmani2020}. In the remainder of this section, we describe the architecture details and training procedure for our method.

%{\noindent \bf Model Architecture:}
\subsection{Model Architecture}
 Our model contains a mixture encoder to transform waveform into encoding, and a backbone to extract source encoding from mixture encoding following \cite{luo2020dual} and \cite{nachmani2020}. Instead of using a single decoder head with a fixed number of output channels, we replaced it with a set of decoder heads, each having a different number of output channels, where one channel contains source from one speaker. We also added a count head that chooses which decoder head to use during inference.

{\noindent \bf Encoder \& Backbone:} 
As in \cite{nachmani2020}, we use convolution with ReLU to encode mixture waveform, then use repeated MulCat blocks as the backbone separation network.

{\noindent \bf Count-Head:}
We train a speaker count head in as an additional branch in parallel with the decoder heads. Given the output tensor from the backbone network, we first linearly transform the feature dimension, then apply global average pooling and ReLU. We then linearly project the result to the set of possible decoder choices, and apply softmax to the output.

{\noindent \bf Decoder-Heads:}
We use a list of decoders, as in \cite{luo2020dual} and \cite{nachmani2020}. For the $k^{\text{th}}$ decoder, given an input tensor with feature dimension $N$, we apply PReLU with a channel-independent parameter, and use 1x1 convolution to project feature dimension to $N\times k$ speakers. We then divide the projected tensor into $k$ tensors, each with feature dimension $N$, and transform the tensor back to audio with overlap-and-add.

%{\noindent \bf Training:}
\subsection{Training}
To train the count-head, we formulate it as a classification task, \ie, we minimize the cross-entropy loss,
\be
\cL_{\text{count}(\rvx, \gY)}  = -\sum_k^K \mathbf{1}_{|\gY|=k} \cdot \log{\hat{p}(|\gY|=k)(\rvx)},
\ee
where $\mathbf{1}$ denotes the indicator function and $\hat{p}(|\gY|=k)$ denotes the predicted probability that the mixed input audio, $\rvx$, has $k$ speakers. 

Next, to train the decoder-heads, we utilize the permutation invariant loss, uPit~\cite{kolbaek2017multitalker}, on the decoder-head selected by the ground-truth number of speakers, \ie, 
\be
\cL_{\text{decoders}}(\rvx, \gY) = \sum_k \mathbf{1}_{|\gY|=k} \cdot \text{uPIT}(\gY, \hat{\gY}_k),
\ee
where $\hat{\gY}_k$ denotes the output from the $k^{\text{th}}$ decoder-head and 
\be
\text{uPIT}(\gY, \hat{\gY_k}) = - \max_\pi \sum_{n}\text{SI-SNR}(\rvy^{\pi(n)}, \hat{\rvy}_k^n),
\ee
where $\pi$ denotes a permutation on the speaker channels, and SI-SNR stands for scale-invariant signal-to-noise ratio, as defined in \cite{1643671}.

Finally, we balance the two losses with a hyper-parameter $\alpha$, and train over a dataset of paired mixed inputs and separated audio, \ie, $\cD = \{(\rvx, \gY)\}$ is as follows,
\be
\min_\theta \sum_{(\rvx, \gY) \in \cD} \alpha \cdot \cL_{\text{count}}(\rvx, \gY) + (1-\alpha) \cdot \cL_{\text{decoders}}(\rvx, \gY).
\ee

\subsection{Inference}
%{\noindent \bf Inference:}
At test time, the ground-truth number of speakers is not available. In this case, we use the estimated number of speakers from the count-head to select which decoder-head to run, therefore,
\be
\hat{\gY} = \hat{\gY}_{\hat{c}}, \;\; \hat{c} = \argmax_k \hat{p}(|\gY|=k)
\ee
is the final prediction given $\rvx$.
%where
%\be
%\hat{c} = \argmax_k \hat{p}(|\gY|=k).
%\ee

%{\noindent \bf Evaluation Metric:}

\begin{table}[t]%[t]
\centering
\begin{tabular}{c c c c c}
\hline
\bf Model & \bf 2 & \bf 3 & \bf 4 & \bf 5\\
\hline
Model-Select(DPRNN)\cite{nachmani2020}* & 81.3 & 64.4 & 46.2 & 85.6\\
Model-Select(Mulcat)\cite{nachmani2020}* & 84.6 & 69.0 & 47.5 & 92.3\\
Attractor Network\cite{xiao2020} & 95.7 & 97.6 & - & -\\
OR-PIT\cite{Takahashi2019RecursiveSS} & \multicolumn{2}{c}{95.7} & - & - \\
Ours & 99.9 & 99.2 & 97.6 & 97.3 \\
\hline

\end{tabular}
\caption{Performance of source counting; Each column is recall for corresponding number of speakers. For OR-PIT, only overall accuracy is provided.}
\label{tab:acc}
\end{table}

\subsection{Evaluation Metric}
Evaluating a system for source separation with variable number of speakers remains an open discussion. It may seem that standard metrics, \eg SI-SNR, are directly applicable, however these metrics require the number of predicted signals and ground-truth signals to be identical. When the system incorrectly predicts the number of speakers, it is unclear how to compute SI-SNR.
%However, there may be cases where the system incorrectly predicts the number of speakers. 

Prior work~\cite{nachmani2020} computes a metric as follows: Let $\hat{S}$ be the number of predicted speaker and $S$ be the ground-truth. In case (a): When $\hat{S} > S$, they compute the correlation between all audio pairs and keep $S$ speakers from the prediction. In case (b): When $\hat{S} < S$, they duplicate $S-\hat{S}$ speakers with the highest correlation to the ground-truth samples. With the speaker number matched, they compute the standard SI-SNR. We note that this choice of duplication / dropping relies on the ground-truth signal. This is not desirable, as a post-processing procedure should not be dependent on the ground-truth.

We believe that it is more natural to add ``silence'' speakers, either ground-truth or the prediction, until the number of speakers between the ground-truth and prediction are identical. Intuitively, a two-speaker mixed signal can be thought of as a three-speaker mixed signal where one of the speakers is silence. However, we run into the issue that SI-SNR is equal to negative infinity if the signal is zero. 

To avoid this, instead of padding with silence, we choose a negative penalty term $\gP_\text{ref}$ that would be defined as the approximation to the SI-SNR measured if padded by silence. We name this metric penalized-SI-SNR (P-SI-SNR). 

Given dataset $\cD = \{(\rvx, \gY)\}$, P-SI-SNR is defined as
\be
\frac{1}{|\cD|}\sum_{(\rvx, \gY) \in \cD} \frac{1}{\text{max}(|\gY|, |\hat{\gY}|)} (\cL_\text{match} + \cL_\text{pad}),
\ee
where $\hat{\gY}=\left\{\rvy^1, ..., \rvy^{\hat{c}}\right\}$, $\hat{c}$ being the number of predicted sources, and $\cL_{\text{match}}$ and $\cL_{\text{pad}}$ are defined as follows:
\be
\begin{aligned}
& \cL_\text{match} =
\max_{\pi}\sum_{n=1}^{\text{min}(|\gY|, |\hat{\gY}|)} \text{SI-SNR}(\rvy^{\pi(n)}, \hat{\rvy}^n)   \\
& \cL_\text{pad} = \gP_\text{ref} \cdot \bigg||\gY| - |\hat{\gY}| \bigg|.
\end{aligned}
\ee

We believe that our proposed metric is intuitive and naturally balances between the reconstruction quality and misclassifications in number of speakers. 
We will next discuss two possible choices of $\gP_\text{ref}$.
%\begin{itemize}
%  \item 

\begin{table}[t]%[t]
\centering
\begin{tabular}{c c c c c}
\hline
\bf Model & \bf 2 & \bf 3 & \bf 4 & \bf 5\\
\hline
Conv-Tasnet\cite{luo2018tasnet}* & 15.3 & 12.7 & - & -\\
DPRNN\cite{luo2020dual}* & 18.8 & - & - & -\\
DPRNN\cite{nachmani2020}* & 18.21 & 14.71 & 10.37 & 8.65\\
Mulcat\cite{nachmani2020}* & 20.12 & 16.85 & 12.88 & 10.56\\
\hline\hline
Attractor Network\cite{xiao2020} & 15.3 & 14.5 & - & -\\
OR-PIT\cite{Takahashi2019RecursiveSS} & 14.8 & 12.6 & 10.2 & - \\
Ours & 19.1 & 14.1 & 9.3 & 5.9 \\
\hline

\end{tabular}
\caption{Oracle SNR; Each column shows results averaged from all mixtures with corresponding number of speakers. *models above double-line are models with fixed number of speakers.}
%only apply to a fixed number of speakers}
\label{tab:oracle-SNR}
\end{table}

{\bf \noindent Measuring $\gP_{\text{ref}}$ from data:}
  One solution is to choose the ``silence'' as some zero-mean noise distribution. In this case, we measure the SNR empirically based on 
  %\ray{working on this...}
  %We can write
  %\be
  %\begin{aligned}
  %\text{SI-SNR}(\hat{\rvy},\rvy_\text{truth})&=\text{SI-SNR%}(\rvy_\text{noise}, \rvy_\text{truth})\\&\equiv %\text{SNR}(\rvy_\text{noise}, \rvy_\text{signal})
  %\end{aligned}
  %\ee
  %The last equivalence follows from the fact that the scaling and dot product computation in the SI-SNR formula
  %are used to decompose $s_\text{target}, e_\text{interf}$ from $\hat{\rvy}$~\cite{1643671}, but those would be unnecessary if we know that the predicted source is pure noise. 
  samples from WSJ0 recordings. 
  We cut out 0.75 second noise segments from their start, repeat those segments to match the length of recordings, and measure the energy ratio between noise files and recordings. Based on the average of our measurements, we set $\gP_\text{ref}$ to be -30dB.

{\bf \noindent Setting $\gP_{\text{ref}}$ as average SI-SNR:}  
  Another intuitive way to penalize SI-SNR is to have 
  %Another intuitive assumption is to have 
  each underestimated or overestimated speaker cancel out the positive contribution to SI-SNR of 
  %from the SI-SNR of 
  a correctly predicted speaker. Therefore, we choose $\gP_\text{ref}$ to be the negative of the average SI-SNR from oracle source counting.
%\end{itemize}

\begin{table*}[t]%[t]
\centering
\begin{tabular}{c c c c c}
\hline
$\gP_\text{ref} = -30dB$ & \bf 2 & \bf 3 & \bf 4 & \bf 5\\
\hline
Model-Select(DPRNN)\cite{nachmani2020}* & 15.2 & 10.7 & 6.0 & 7.7\\
Model-Select(Mulcat)\cite{nachmani2020}* & 17.5 & 13.21 & 8.4 & 10.0\\
\hline
Attractor Network\cite{xiao2020} & 14.7 & 14.2 & - & -\\
OR-PIT\cite{Takahashi2019RecursiveSS} & \multicolumn{2}{c}{13.1} & - & - \\
Ours & 19.1 & 14.0 & 9.2 & 5.8 \\
\hline
\end{tabular}
\quad
\begin{tabular}{c c c c c}
\hline
$\gP_\text{ref} = -\text{SI-SNR}_\text{oracle}$ & \bf 2 & \bf 3 & \bf 4 & \bf 5\\
\hline
Model-Select(DPRNN)\cite{nachmani2020}* & 15.9 & 12.1 & 8.1 & 8.2\\
Model-Select(Mulcat)\cite{nachmani2020}* & 18.1 & 14.2 & 10.2 & 10.3\\
\hline
Attractor Network\cite{xiao2020} & 14.9 & 14.3 & - & -\\
OR-PIT\cite{Takahashi2019RecursiveSS} & \multicolumn{2}{c}{13.4} & - & - \\
Ours & 19.1 & 14.0 & 9.3 & 5.9 \\
\hline
\end{tabular}

\caption{P-SI-SNR of each model; For OR-PIT, result is computed by averaging the P-SI-SNR for both 2 and 3 speakers computed with 95.7\% recall. Note that models with lower max speaker count generally have higher accuracy, since fewer classes implies a higher P-SI-SNR.}
\label{tab:PI-SI-SNR}
\end{table*}
%We choose the penalty to be -30dB.

%\input{drafts/app1}
%!TEX root = ../main.tex
\section{Experiments}
We first describe our implementation details for dataset preparation, training, testing, model architecture. Next, we provide quantitative comparisons with baselines and demonstrate that our approach achieves state-of-the-art performance in source counting and comparable performance in source separation.

%\ray{Some intro here.}
%\subsection{Datasets}
{\noindent \bf Datasets:}
We train on WSJ0-2mix and WSJ0-3mix~\cite{hershey2016deep},
in addition to WSJ0-4mix and WSJ0-5mix\cite{nachmani2020}. We take 4-second chunks of all audios with 2-second overlap, and pad any chunks at the end that are above 2 seconds. We remove all mixtures below 2s. As mixtures have length equal to the shortest source, those with more speakers are shorter and have less chunks. In our training set, 2, 3, 4, 5 speakers all have 20000 audios, and respectively have [24773, 19066, 15986, 13809] chunks.

{\noindent \bf Training Procedure:}
%We re-sample the chunks to ensure mixtures with every speaker count are visited equally. 
For each epoch, we use weighted re-sampling with replacement to ensure that chunks for each speaker number are sampled with equal probability. We set probability of choice for each chunk inversely proportional to number of chunks with the same speaker count. %An epoch length is to be the number of total chunks in our training set.
We train our model using Adam~\cite{kingma2014adam} with learning rate $5e-4$,  decay of $0.94$ every epoch, and batch size of $4$. %We set $\alpha$ to 0.035.
In total, we train our model for 40 epochs, which is much less than the 100 epochs in most previous papers~\cite{Takahashi2019RecursiveSS,luo2020dual}. 

{\noindent \bf Testing Procedure:}
%During testing, we compute P-SI-SNR on the entire signal. %Since the network is trained on chunks, we also perform prediction on chunks.
Given an audio signal, We first transform the full audio into chunks. We use the count head to predict which decoder head to use for each chunk and select the decoder head with the highest votes. Using the selected decoder head, we compute separated sources for each chunk, and use the overlap regions to reorder the predicted source channels in a streaming fashion. Lastly, we use overlap-and-add to recover predicted sources for the full audio, and remove the padding at the end chunk. %iWe then compute the P-SI-SNR for both choices of $\gP_\text{ref}$ using the formulas above.

{\noindent \bf Architecture Details}:
 For the encoder, we use a convolution kernel size of 8, stride 4, and 256 feature channels. For backbone, we use LSTM with hidden layer size 256. Similar to \cite{nachmani2020}, we use multi-stage loss, but do not use speaker ID loss for simplicity. During training, we train both the decoders and count-heads with multi-loss, with one set of output after each pair of Mulcat blocks.\footnote{See project page for more details: \url{https://junzhejosephzhu.github.io/Multi-Decoder-DPRNN/}} %During testing, we only compute the count-head and decoder outputs after the final pair of Mulcat blocks.

{\noindent \bf Comparisons with Baselines:}
Many of the systems for variable speaker source separation are not publicly available, therefore we cannot directly compare with them on our proposed P-SI-SNR. 
%We compute the baselines using P-SI-SNR based on provided statistics from various previous systems. As the systems are not publicly available, 
To compare, we use the reported numbers from their paper. Note that since we do not have the exact SI-SNR, in the case of speaker mismatch, we compute an upper bound for the models using their published statistics on oracle SI-SNR and speaker counting accuracy.

%
%{\noindent \bf Computing Upper bound for Baselines:}
For computing this upper bound, we assume that each misclassification of speaker number is an overestimate by one, and all the other channels have oracle SI-SNR. This is an upper bound because oracle SNR is always higher than non-oracle SNR, and the ratio of (contribution from correct channels)/penalty is greatest if the error is an overestimate by one channel.
For a model with $k$ speakers with oracle SI-SNR $x$ and accuracy $a$, the upper bound for P-SI-SNR can be computed as
\be
\mbox{P-SI-SNR} \le a \times x + (1 - a) \times \frac{(k \times x + \gP_\text{ref})}{k + 1}
\ee
%\ray{Not sure if the reader can understand this.}
%For example, for Attractor Network, 2-speakers,
%\be
%\begin{aligned}
%\text{P-SI-SNR} = & 0.957 \times 15.3 + (1 - 0.957) %\times \frac{(2 \times 15.3 - 30)}{3} \\= & 14.7
%\end{aligned}
%\ee

{\noindent \bf Quantitative Results:}
We report quantitative comparisons for source counting performance in ~\tabref{tab:acc}, oracle SNR in~\tabref{tab:oracle-SNR}, and our proposed P-SI-SNR in~\tabref{tab:PI-SI-SNR}. We note that models with * are not directly comparable to our model as they train a model \textit{for each} speaker number, where we have a \textit{single model} for all speakers.
%We do not highlight anything in our results table, because 1. only Attractor Network, OR-PIT, and our model train a single network on mixtures of variable speaker counts, while Conv-Tasnet, DPRNN, and Mulcat train one network for each source count, and therefore do not generalize to more speakers. 2. It is unfair to compare the recall between models with different number of output classes.

As can be seen from~\tabref{tab:acc}, in the source counting task, our model outperforms all other models, even those with fewer possible choices of speaker counts. Our approach remains competitive in source separation when evaluated using Oracle-SNR, see~\tabref{tab:oracle-SNR}.
Lastly, in~\tabref{tab:PI-SI-SNR}, when $\gP_\text{ref}$ is set to -30 dB, our P-SI-SNR also outperforms all other models in 2-speaker and 4-speaker cases, and achieves similar results to best model in the 3-speaker case.

%!TEX root = ../main.tex

\section{Conclusion}
We present a unified approach to single channel speech separation with an unknown number of speakers. With our proposed multi-decoder architecture and count-head, our model requires a single forward-pass at test-time on a single network. In our experiments, we demonstrate that our model achieves state-of-the-part performance in source counting and competitive source separation quality. Additionally, we propose a new evaluation metric for evaluating source separation with an unknown number of speakers, in which we penalize SI-SNR when the number of sources estimated is incorrect.

\clearpage
{
\bibliographystyle{IEEEbib}
\bibliography{sections/refs}
}

\end{document}